\documentstyle[12pt]{article}

\makeatletter
\renewcommand\thesubsection{\thesection.\@arabic\c@subsection}
\makeatother

\newcommand{\sect}[1]{\setcounter{equation}{0}\section{#1}}

\newcommand {\beq}{\begin{equation}}
\newcommand {\eeq}{\end{equation}}
\newcommand {\beqa}{\begin{eqnarray}}
\newcommand {\eeqa}{\end{eqnarray}}         %Equation version
\newcommand {\beqs}{\begin{eqnarray*}}
\newcommand {\eeqs}{\end{eqnarray*}}
\newcommand {\bds}{\begin{displaymath}}
\newcommand {\eds}{\end{displaymath}}
\newcommand {\n}{\nonumber\\}

\newcommand {\bebb}{\begin{thebibliography}}
\newcommand {\eebb}{\end{thebibliography}}      %Reference version
\newcommand {\bbit}{\bibitem}
\def\a{\alpha}

\def\k{\kappa}

\def\pd{\prod}

\def\ra{\rangle}

\def\dg{\dagger}

\def\ph{\phi}

\def\journal#1&#2(#3){\unskip, \sl #1\ \bf #2 \rm(19#3) }
\def\andjournal#1&#2(#3){\sl #1~\bf #2 \rm (19#3) }

\def\dz{\frac{d}{dz}}

\begin{document}

%\begin{titlepage}

\begin{flushright}
\end{flushright}

\vskip 1cm

\begin{center}
%\title
{\Large\bf Polynomial algebras and exact solutions of general quantum non-linear optical models II:
Multi-mode boson systems}

\vspace{1cm}

%\author{
{\large Yuan-Harng Lee $^a$, Wen-Li Yang $^{b}$ and Yao-Zhong Zhang $^{a}$}
\vskip.1in

$a.$ {\em School of Mathematics and Physics,
The University of Queensland, Brisbane, Qld 4072, Australia}

$b.$ {\em Institute of Modern Physics, Northwest University,
Xi'an 710069, China}

\end{center}

\date{}

%\maketitle

%\vspace{2cm}

\begin{abstract}
We present higher order polynomial algebras which are the dynamical symmetry algebras of
a wide class of multi-mode boson systems in non-linear optics and laser physics.  
We construct their unitary representations and the corresponding single-variable
differential operator realizations. We then use the results to obtain exact
(Bethe ansatz) solutions to the multi-mode boson systems, including the Bose-Einstein Condensate 
models as the special cases. 
%We also establish the spectral equivalence
%between the BEC models and certain quasi-exactly solvable Sch\"ordinger potentials.

\end{abstract}

\vskip.1in



%\vspace{0.5cm}

%\end{titlepage}

\setcounter{section}{0}
\setcounter{equation}{0}

\sect{Introduction }

This paper is the second of a series of three devoting to polynomial algebraic structures and exact solutions of
general quantum non-linear optical models. In the previous work \cite{Yuan10}, we exactly solved  
a class of two-mode boson systems by identifying and applying the underlying dynamical polynomial 
algebra symmetries. In this article, we generalize the results of \cite{Yuan10} and deal with a general class of 
multi-mode boson systems.  
 
There has been constant interest in the study of deformed Lie algebras and their applications to physics. 
This has been motivated somewhat by the realization that the traditional linear Lie algebras and their affine 
counterparts may be too restrictive as there is no reason for symmetries 
to be linear in the first place. 

Polynomial algebras are a particular type of non-linear deformations of Lie algebras \cite{Smith90}. 
Their applications span across a diverse field of theoretical physics whereby
they appear in topics such as quantum mechanics, Yang-Mills type gauge theories,  quantum non-linear optics,
integrable systems and (quasi-)exactly solvable models, to name a few (see e.g. \cite{Higgs79}-\cite{Klishevich01}).
Different realizations of polynomial algebras have been extensively studied in \cite{Karassiov94,Karassiov92-00,Karassiov02}
(see also \cite{Zhedanov92,Kumar01,Beckers99,Debergh00} for 
%A number of studies have been undertaken to construct unitary representations of these algebras.
%For instance, 
differential operator realizations of certain quadratic and cubic algebras
%have been explored in \cite{Zhedanov92,Kumar01} and also in \cite{Beckers99,Debergh00} 
in connection with quasi-exact integrability \cite{Turbiner88,Ushveridze94,Gonzarez93}).

One of the main results of this paper is the derivation of the exact eigenfunctions and energy
eigenvalues of the multi-boson Hamiltonian,
\beqa
H&=&\sum_i^{r+s}w_iN_i+ \sum_{i \le j}^{r+s} w_{ij}N_iN_j\n
& & + g\left( a_1^{\dagger k_1}\cdots a_r^{\dagger k_r}
a_{r+1}^{k_{r+1}}\cdots a_{r+s}^{k_{r+s}} + a_{1}^{k_{1}}\cdots a_{r}^{k_{r}}a_{r+1}^{\dagger k_{r+1}}
\cdots a_{r+s}^{\dagger k_{r+s}} \right),  \label{bosonH1}
\eeqa
where and throughout $a_i~(a_i^\dagger)$ and $N_i=a^\dagger_ia_i$
are bosonic annihilation (creation) and number operators,
respectively, and $w_i,~w_{ij}$ and $g$ are real coupling constants. 
Hamiltonians of the form (\ref{bosonH1}) appear in the description of various physical systems of interest
such as non-linear optics and laser physics. % and BECs \cite{Anderson95}-\cite{Donley02}. 
Using the Bargmann representation, it was shown in \cite{Alvarez02} that the multi-boson system is  
quasi-exactly solvable (see also \cite{Alvarez95}). Some special cases of the system have been studied in
e.g. \cite{Andreev02,Links03} by means of the Algebraic Bethe Ansatz method. Here we determine the underlying 
dynamical polynomial algebra symmetries and the exact solutions of the general Hamiltonian (\ref{bosonH1}). 
The polynomial algebras and representations constructed 
in this paper are of interest in their own right, in view of their potential applications in other fields.

This paper is organized as follows. In sections 2 we introduce higher order polynomial algebras with
multi-mode boson realizations. We construct their unitary representations in the Fock space 
and the corresponding single-variable 
differential operator realizations. We then identify the polynomial algebras as the dynamical symmetries of 
Hamiltonian (\ref{bosonH1}) in section 3, and solve for the eigenvalue problem via the Functional Bethe 
Ansatz method (see e.g. \cite{Wiegmann94,Sasaki09,Sasaki01}). In section 4,
we present explicit results for models corresponding to the special cases $r+s \le 4$, providing a unified treatment
of the so-called Bose-Einstein Condensate (BEC) models.
%In section 6, we establish the spectral correspondence of these specific models with quasi-exactly solvable (QES)
%Schr\"odinger potentials. 
We summarize our results in section 5 and discuss further avenues of investigation.

%\setcounter{section}{1}
%\setcounter{equation}{0}

%%%%%%%%%%%%%%%%%%%%%%%%%%%%%%%%%%%%%%%%%%%%%%%%%%%%%%%%%%%%%%%%%%%%%%%%%%%%%%%%%%%%%%%%%%%%%%%%%%%%%%%%%%%%%%%%%%%%%%%%%%%%%%%

%\sect{Polynomial deformations of the oscillator algebra}
\sect{ Polynomial deformations of $sl(2)$ algebra}

We first briefly review the polynomial algebras defined in \cite{Yuan10}. Let $k$ be a positive integer, 
$k=1,2,\cdots$. Consider the polynomial algebra of degree $k-1$ defined by %the commutation relations
\beqa
[ Q_0,Q_{\pm} ] &=& \pm Q_{\pm}, \n
\left[ Q_+,Q_{-} \right]  &=& \ph^{(k)}(Q_0)-\ph^{(k)}(Q_0-1), \label{su11-poly-alg}
\eeqa
where
\beq
\ph^{(k)}(Q_0)= - \pd_{i=1}^{k} \left( Q_0+\frac{ik-1}{k^2} \right)
+\pd_{i=1}^k \left(\frac{i-k}{k} - \frac{1}{k^2} \right)
\eeq
is a polynomial in $Q_0$ of degree $k$. The algebra admits Casimir operator of the form
\beq
C=Q_-Q_+ +\ph^{(k)}(Q_0)=Q_+Q_-+\ph^{(k)}(Q_0-1).\label{casimir}
\eeq
%For $k=1$, algebra (\ref{su11-poly-alg}) reduces to the oscillator algebra and thus 
%can be viewed as polynomial deformations of the oscillator algebra. 
%
As was shown in \cite{Yuan10}, algebra (\ref{su11-poly-alg}) has an infinite dimensional irreducible unitary 
representation given by the following one-mode boson realization,
\beq
Q_+=\frac{1}{(\sqrt{k})^k}(a^{\dg})^k ,~~~~~
Q_-=\frac{1}{(\sqrt{k})^k}(a)^k , ~~~~~
Q_0 =\frac{1}{k} \left(a^{\dg}a+ \frac{1}{k} \right).\label{su11-poly-boson}
\eeq
In this realization, the Casimir (\ref{casimir}) takes fixed value 
\beq
C=\prod_{j=1}^k\left(\frac{j-k}{k}-\frac{1}{k^2}\right),
\eeq
which corresponds to the common eigenvalue  \cite{Yuan10} %$-\pd_i^{k}\left(q+\frac{i-k}{k}-\frac{1}{k^2}\right) 
%+ \pd_{i=1}^k\left(\frac{i-k}{k} -\frac{1}{k^2} \right)$ 
for the direct sum of $k$  irreducible representations with
%It was shown \cite{Yuan10} that the irreducible unitary representations of (\ref{su11-poly-alg}) with the above boson realization are labeled by 
quantum number $q=\frac{1}{k^2},~\frac{k+1}{k^2},~\cdots,~ \frac{(k-1)k+1}{k^2}$ in each case. Note that  
for $k=2$, $C$ becomes $3/16$ and $q$ reduces to the well-known quantum number $K=1/4, 3/4$ used in $su(1,1)$.  
Therefore quantum number $q$ is a very natural choice for labeling the Fock states, 
denoted as $|q,n \ra $, of the irreducible representations of (\ref{su11-poly-alg}). Explicitly, 
\beqa 
|q,n \ra = {a^{\dg k(n+q-\frac{1}{k^2})} \over \sqrt{[k(n+q-{1\over k^2})]!}}|0 \ra, ~~~~ %\label{lw-state2} \n 
n = 0,1,\cdots.
%, \hspace{1cm} q=\frac{1}{k^2},~\frac{k+1}{k^2},~\cdots,~ \frac{(k-1)k+1}{k^2}
\label{su11-rep}
\eeqa 
The action of $Q_0, Q_{\pm}$ on these states are as follows
\beqa
Q_0|q,n\ra &=& (q+n)|q,n\ra, \n
Q_+|q,n\ra &=& \pd_{i=1}^{k}\left( n+q+ \frac{i k-1}{k^2}\right)^\frac{1}{2}\;|q,n+1\ra, \n
Q_-|q,n\ra &=& \pd_{i=1}^{k}\left( n+q -\frac{(i-1)k+1}{k^2}\right)^\frac{1}{2}\;|q,n-1\ra. %, \n
%C|q,n \ra  &=& \left[\pd_{i=1}^{k}\left(\frac{i-k}{k}- \frac{1}{k^2}\right)\right]|q,n \ra. 
\eeqa

Now let us take $r$ mutually commuting copies of algebra (\ref{su11-poly-alg}), 
$\{Q_{+}^{(i)},Q_{-}^{(i)},Q_{0}^{(i)} \}$, 
of degree $k_i-1$,  with the multi-boson realization ($i=1, \cdots, r$)
\beq
Q_+^{(i)}=\frac{1}{(\sqrt{k_i})^{k_i}}(a_i^{\dg})^{k_i} ,~~~~~
Q_-^{(i)}=\frac{1}{(\sqrt{k_i})^{k_i}}(a_i)^{k_i} , ~~~~~
Q_0^{(i)} =\frac{1}{k_i} \left(a_i^{\dg}a_i+ \frac{1}{k_i} \right).
\eeq
Consider the algebra generated by 
\beq
{\cal Q}_+ = Q_+^{(1)}\cdots Q_+^{(r)},~~~~
{\cal Q}_- = Q_-^{(1)}\cdots Q_-^{(r)},~~~~
{\cal Q}_0 = \frac{1}{r}\,\sum_{i=1}^r Q_0^{(i)}.\label{su11-poly2-alg}
\eeq 
It can be shown that ${\cal Q}_{0,\pm}$ form the polynomial algebra of degree ($\sum_{i=1}^r k_i-1$),
\beqa 
\left[ {\cal Q}_0,{\cal Q}_{\pm} \right] &=& \pm {\cal Q}_{\pm}, \n 
\left[{\cal Q}_+,{\cal Q}_- \right] &=& -\prod_{i=1}^r \phi^{(k_i)}({\cal Q}_0,\{{\cal L}\})
   +\prod_{i=1}^r \phi^{(k_i)}({\cal Q}_0-1, \{{\cal L}\}).\label{su11-poly2-alg2}
\eeqa 
In the above,  
%$C_{i},\,i=1,2,\cdots, M$ are the Casimirs of the $M$ mutually commuting algebras,
%\beq
%C_{i}=Q^{(i)}_-Q_+^{(i)} +\ph^{(k_i)}(Q_0^{(i)})=Q_+^{(i)}Q_-^{(i)}+\ph^{(k_i)}(Q_0^{(i)}-1),\label{casimir-Ci}
%\eeq
$\{{\cal L}\}$ stands for the set $\{{\cal L}_1,\cdots, {\cal L}_{r-1}\}$, with
\beqa 
 {\cal L}_j &=&Q_0^{(j)}-Q_0^{(j+1)}  \label{su11-poly2-central}
\eeqa 
being the $r-1$ central elements of algebra (\ref{su11-poly2-alg2}), $[{\cal L}_j, {\cal Q}_{0,\pm}]=0$, and
\beqa 
 \phi^{(k_i)}({\cal Q}_0,\{{\cal L}\}) = \prod_{j=1}^{k_i}\left({\cal Q}_0+ \sum_{\mu=i}^{r-1} {\cal L}_\mu
- \frac{1}{r}\sum_{\mu=1}^{r-1}\mu {\cal L}_\mu 
  + \frac{jk_i-1}{k_i^2} \right) % + \prod_{j=1}^{k_i}\left(\frac{j-k_i}{k_i} - \frac{1}{k_i^2} \right)\n
\eeqa 
is a polynomial in ${\cal Q}_0$ and ${\cal L}_i$ of degree $k_i$. From (\ref{su11-poly2-central}), we can show that 
\beq
Q_0^{(i)}=\sum_{j=i}^{r-1}{\cal L}_j+Q_0^{(r)},~~~~~~Q_0^{(r)}={\cal Q}_0-\frac{1}{r}\sum_{j=1}^{r-1}j{\cal L}_j.\label{QL-relations}
\eeq

%Some remarks is in order. For $M=2, k_1=k_2=1$, algebra (\ref{su11-poly2-alg2}) reduces to the (linear) $su(1,1)$ algebra.
%In general, the right hand side of (\ref{su11-poly2-alg2}) depends on the Casimir operators $C_i$ and thus 
%the polynomial algebra is Casimir dependent. In other words, the polynomial algebra structure is realization or representation dependent:
%different realizations (or representations) satisfy different polynomial algebras (corresponding to different 
%values of Casimirs). This is one of the interesting characteristics of a polynomial algebra in general. 
  
%It can be easily proved that the following multi-boson operators
%\beq 
%{\cal Q}_+ = \prod_{i=1}^M \frac{(a_i^{\dg})^{k_i}}{(\sqrt{k_i})^{k_i}},~~~~~
%% \cdots  \frac{(a_M^{\dg})^{k_M}}{\sqrt{k_M}^{k_M}}, \n
%{\cal Q}_- = \prod_{i=1}^M \frac{(a_i)^{k_i}}{(\sqrt{k_i})^{k_i}},~~~~~
%% \cdots  \frac{(a_M)^{k_M}}{\sqrt{k_M}^{k_M}}, \n
%{\cal Q}_0 = \frac{1}{M}\,\sum_{i=1}^M\frac{1}{k_i}\left(a_i^{\dg}a_i + \frac{1}{k_i} \right)
%\eeq 
%give a realization of the polynomial algebra (\ref{su11-poly2-alg2}) with the Casimirs $C_{i}, \;i=1,2,\cdots, M,$ 
%taking the particular values
%\beq
%C_{i}=\prod_{j=1}^{k_i}\left(\frac{j-k_i}{k_i}-\frac{1}{k_i^2}\right). \label{Ci-values}
%\eeq
%This realization provides an infinite dimensional representation of the algebra. 
Unitary irreducible representations of polynomial algebra (\ref{su11-poly2-alg2}) are infinite dimensional.
The corresponding Fock states take the form
$\prod_{i=1}^r|q_i,n_i \ra $, where $n_i =0,1,\cdots,$ and $q_i = \frac{1}{k_i^2},\frac{k_i+1}{k_i^2},\cdots, \frac{(k_i-1)k_i+1}{k_i^2}$.
%\beqa  
%\prod_{i=1}^M|q_i,n_i \ra &=& \prod_{i=1}^M\frac{a_i^{\dg k_i(n_i+q_i-\frac{1}{k_i^2})}}
%{\sqrt {\left[k_i(n_i+q_i-\frac{1}{k_i^2})\right]!}} |0\ra, \n 
%n_i &=&0,1,\cdots,~~~~ ~q_i = \frac{1}{k_i^2},\frac{k_i+1}{k_i^2},\cdots, \frac{(k_i-1)k_i+1}{k_i^2}.
%\eeqa  
By means of the central elements (\ref{su11-poly2-central}), we can show that the Fock states become 
\beqa 
|\{q\},n,\{l\} \ra = \prod_{i=1}^r|q_i,s_i+q_r-q_i+n \ra  
= \prod_{i=1}^r \frac{a_i^{\dg k_i(n+q_r+s_i-\frac{1}{k_i^2})}}{\sqrt{[k_i (n+q_r+s_i-\frac{1}{k_i^2})]!}}, 
\eeqa 
where $n=0,1, \cdots$,  and $s_i = \sum_{j=i}^{r-1}l_j$  for $i=1,2,\cdots, r-1$ and  $s_r\equiv 0$;
$\{q\}\equiv \{q_1,\cdots, q_r\}$ and $\{l\}\equiv \{l_1, \cdots, l_{r-1} \} $ 
are the constant values taken by the central elements ${\cal L}_i$.
Note that  $s_i+q_r-q_i\geq 0$ (so $q_r+s_i-\frac{1}{k_i^2}\geq 0$).  
%The actions of (\ref{su11-poly2-alg}) on the Fock states are given by 
%\beqa
%{\cal Q}_0|\{q\},n,\{l\} \ra  &=& \left( q_M+n+\frac{1}{M}\sum_{j=1}^M s_j \right) |\{q\},n,\{l\} \ra ,\n 
%{\cal Q}_+|\{q\},n,\{l\} \ra &=& \left[\prod_{i=1}^M \prod_{j=1}^{k_i}\left(n+q_M+s_i
% +\frac{jk_i-1}{k_i^2}\right)^{\frac{1}{2}} \right]|\{q\},n+1,\{l\} \ra ,  \n 
%{\cal Q}_-|\{q\},n,\{l\} \ra &=&\left[\prod_{i=1}^M\prod_{j=1}^{k_i}\left(n+q_M+s_i-\frac{(j-1)k_i+1}{k_i^2} \right)^{\frac{1}{2}}
% \right]|\{q\},n-1,\{l\} \ra . %, \n
%%n &=& 0,1,\cdots, ~~~~~ q_M = \frac{1}{k_M^2},\frac{k_M+1}{k_M^2}, \cdots, \frac{(k_M-1)k_M+1}{k_M^2}. 
%\eeqa 

%\sect{ Polynomial deformations of $su(2)$ algebra}

%Unitary representations of the polynomial algebra (\ref{su11-poly2-alg2}) are all infinite dimensional. %In this section, 
We now construct polynomial algebra which has finite dimensional unitary irreducible representations and thus can be identified as the 
dynamical symmetry algebra of the multi-boson Hamiltonian (\ref{bosonH1}). 
We do so by considering two mutually commuting algebras % of the preceding section, 
$\{ {\cal Q}_{0,\pm}^{(1)} \}$ of degree $\sum_{i=1}^{r} k_i -1 $ 
and $\{ {\cal Q}_{0,\pm}^{(2)}  \}$ of degree $\sum_{i=r+1}^{r+s} k_i-1 $, with
\beqa
&&{\cal Q}_+^{(1)} =\prod_{i=1}^{r}\frac{(a_i^{\dg})^{k_i}}{(\sqrt{k_i})^{k_i}} , \hspace{5mm} 
{\cal Q}_-^{(1)} = \prod_{i=1}^{r}\frac{(a_i)^{k_i}}{(\sqrt{k_i})^{k_i}} ,
\hspace{5mm}  {\cal Q}_0^{(1)} =\frac{1}{r}\sum_{i=1}^{r}\frac{1}{k_i} \left( N_i+ \frac{1}{k_i} \right), \n
&&{\cal Q}_+^{(2)} =\prod_{i=r+1}^{r+s}\frac{(a_i^{\dg})^{k_i}}{(\sqrt{k_i})^{k_i}} , \hspace{5mm} 
{\cal Q}_-^{(2)} = \prod_{i=r+1}^{r+s}\frac{(a_i)^{k_i} }{(\sqrt{k_i})^{k_i}},
\hspace{5mm}  {\cal Q}_0^{(2)} =\frac{1}{s}\sum_{i=r+1}^{r+s}\frac{1}{k_i} \left( N_i+ \frac{1}{k_i} \right).\n
 \label{su2-poly-boson}
\eeqa
Introducing new generators,
\beq
{\cal{P}}_+ = {\cal Q}_+^{(1)} {\cal Q}_-^{(2)}, ~~~~
{\cal{P}}_- =  {\cal Q}_+^{(2)} {\cal Q}_-^{(1)}, ~~~~
{\cal{P}}_0 = \frac{1}{2}\left(  {\cal Q}_0^{(1)}-  {\cal Q}_0^{(2)} \right). \label{jordan-schwinger}
\eeq
We can easily show that ${\cal P}_{0,\pm}$ form a polynomial algebra of degree $(\sum_{i=1}^{M_1+M_2}k_i-1)$ which
close under the following commutation relations:
\beqa
\left[{\cal{P}}_0,{\cal{P}}_{\pm}\right] &=& \pm {\cal P}_{\pm}, \n
\left[{\cal{P}}_+,{\cal{P}}_- \right] &=& -\prod_{i=1}^{r}{\cal \varphi}_1^{(k_i)}({\cal{P}}_0,{\cal{K}},\{{\cal L}\}^{(1)})
  \prod_{i=r+1}^{r+s}{\cal \varphi}_2^{(k_i)}({\cal{P}}_0,{\cal{K}}, \{{\cal L}\}^{(2)})\n
& &+\prod_{i=1}^{r}{\cal \varphi}_1^{(k_i)}({\cal{P}}_0-1,{\cal{K}}, \{{\cal L}\}^{(1)})
  \prod_{i=r+1}^{r+s}{\cal \varphi}_2^{(k_i)}({\cal{P}}_0-1,{\cal{K}}, \{{\cal L}\}^{(2)}),
 \label{su2-poly}
\eeqa
where $\{{\cal L}\}^{(1)}\equiv\{{\cal L}_1,\cdots,{\cal L}_{r-1}\}$ 
and $\{{\cal L}\}^{(2)}\equiv\{{\cal L}_{r+1},\cdots,{\cal L}_{r+s-1}\}$ with ${\cal L}_i$ 
given by (\ref{su11-poly2-central}) are the central elements of (\ref{su2-poly}), i.e. $[{\cal L}_j, {\cal P}_{0,\pm}]=0$, 
%$\{C\}^{(1)}=\{C_1,\cdots, C_{M_1}\}$ and
%$\{C\}^{(2)}=\{C_{M_1+1},\cdots, C_{M_1+M_2}\}$ with $C_i$ given by (\ref{Ci-values}) satisfy $[C_i, {\cal P}_{0,\pm}]=0$,
\beq
{\cal K} =\frac{1}{2}\left({\cal Q}_0^{(1)}+{\cal Q}_0^{(2)} \right)
\eeq
is the $(r+s-1)$-the central element of the algebra, $\left[ {\cal K} , {\cal {P}}_{0,\pm}\right] = 0$, and finally
\beqa 
&&{\cal \varphi}_1^{(k_i)}({\cal{P}}_0, {\cal K}, \{{\cal L}\}^{(1)}) 
% C_{i}-\prod_{j=1}^{k_i}\left(\frac{j-k_i}{k_i}-\frac{1}{k_i^2}\right)\n
 =\prod_{j=1}^{k_i}\left({\cal K}+{\cal P}_0+ \sum_{\mu=i}^{r-1}{\cal L}_\mu -\frac{1}{r}
  \sum_{\mu=1}^{r-1}\mu{\cal L}_\mu+\frac{jk_i-1}{k_i^2}\right),  \n 
&&{\cal \varphi}_2^{(k_i)}({\cal{P}}_0, {\cal K},  \{{\cal L}\}^{(2)}) 
% C_{i}-\prod_{j=1}^{k_i}\left(\frac{j-k_i}{k_i}-\frac{1}{k_i^2}\right)\n
=\prod_{j=1}^{k_i}\left({\cal K}-{\cal P}_0-1
  +\sum_{\mu=i}^{r+s-1}{\cal L}_\mu-\frac{1}{s}\sum_{\mu=r+1}^{r+s-1}\mu{\cal L}_\mu+\frac{jk_i-1}{k_i^2} \right).\n
\eeqa 
%is a polynomial in ${\cal P}_0$ and $\cal K$ of degree $\sum_{i=1}^{M_1+M_2}k_i$. 
%For $M_1=M_2=1$ and $k_1=k_2=1$, algebra (\ref{su2-poly}) reduces to the (linear) $su(2)$ algebra and thus can be regarded as
%polynomial deformations of $su(2)$.

%In terms of $M_1+M_2$ sets of mutually commuting boson operators acting on
%the tensor product of the Fock spaces, we have the realization
%
%These multi-boson operators satisfy the polynomial algebra (\ref{su2-poly}) with the Casimirs $C_{i}$ taking
%the particular values (\ref{Ci-values}). 
%%\beq
%%C^{(i)}=\prod_{j=1}^{k_i}\left(\frac{j-k_i}{k_i}-\frac{1}{k_i^2}\right),~~~~i=1,2,\cdots,M_1+M_2.
%%\eeq
%This realization gives rise to finite dimensional representations of (\ref{su2-poly}).
Unitary irreducible representations of polynomial algebra (\ref{su2-poly}) are finite dimensional.
To show this, let $|\{q \}^{(1)}, n^{(1)},\{l \}^{(1)} \ra$ and $|\{q\}^{(2)}, n^{(2)},\{l \}^{(2)} \ra $ 
be the Fock states of the algebras $\left\{{\cal Q}_{0,\pm}^{(1)} \right\}$ and $\left\{ {\cal Q}_{0,\pm}^{(2)} \right \}$ 
respectively, where $n^{(1)},~n^{(2)}=0,1,\cdots$, and
%\beqa
$\{q\}^{(1)} \equiv \{q_1, \cdots , q_{r} \}$, %~~~~
$\{q\}^{(2)} \equiv \{q_{r+1}, \cdots ,q_{r+s} \}$. % \n 
%q_i &=& \frac{1}{k_i^2}, \frac{k_i+1}{k_i^2}, \cdots , \frac{(k_i-1)k_i+1}{k_i^2}.
%\eeqa  
The representations of $\left\{ {\cal P}_{0,\pm} \right\} $ %corresponding to the realization (\ref{su2-poly-boson}) 
are then given by the Fock states 
$|\{q \}^{(1)}, n^{(1)},\{l \}^{(1)} \ra \;|\{q \}^{(2)}, n^{(2)},\{l \}^{(2)} \ra $.
Since $\cal K$ is a central element of the algebra, it must be a constant, denoted as $\k$ below,
on any irreducible representations. This imposes the following constraint %on the values of $n^{(1)}$ and $n^{(2)}$,
%\beqa
%&& {\cal K} |\{q \}^{(1)}, n^{(1)},\{l \}^{(1)} \ra |\{q \}^{(2)}, n^{(2)},\{l \}^{(2)} \ra  \n 
%&& = \frac{1}{2}\left(q_{M_1}+n^{(1)}+\frac{1}{M_1}\sum_{i=1}^{M_1}s_i^{(1)} +q_{M_1+M_2}+n^{(2)}+\frac{1}{M_2} \sum_{i=M_1+1}^{M_1+M_2}s_i^{(2)} \right) \n 
%&& 	\times |\{q \}^{(1)}, n^{(1)},\{l \}^{(1)} \ra |\{q \}^{(2)}, n^{(2)},\{l \}^{(2)} \ra \n 
%&& =\k|\{q \}^{(1)}, n^{(1)},\{l \}^{(1)} \ra |\{q \}^{(2)}, n^{(2)},\{l \}^{(2)} \ra 
%\eeqa
\beq
n^{(1)}+n^{(2)}=2\k-q_{r}-q_{r+s}-t,
%-\left(q_{M_1}+\frac{1}{M_1}\sum_{i=1}^{M_1}s_i^{(1)} +q_{M_1+M_2}+\frac{1}{M_2}\sum_{i=M_1+1}^{M_1+M_2}s_i^{(2)} \right),
\eeq
where 
\beqa 
t&=&\frac{1}{r}\sum_{i=1}^{r}s_i^{(1)}+\frac{1}{s}\sum_{i=r+1}^{r+s}s_i^{(2)},\n
s_i^{(1)} &=& \sum_{j=i}^{r-1}l_j ,~~i= 1, \cdots, r-1,~~~ s_{r}^{(1)}=0,\n  	
s_i^{(2)} &=& \sum_{j=i}^{r+s-1}l_j ,~~ i=r+1, \cdots, r+s-1 , ~~~ s^{(2)}_{r+s}=0.
\eeqa 
%Let 
%\beq 
%t=\frac{1}{M_1}\sum_{i=1}^{M_1}s_i^{(1)}+\frac{1}{M_2}\sum_{i=M_1+1}^{M_1+M_2}s_i^{(2)}.
%\eeq 
Let ${\cal N}=2\k-q_{r}-q_{r+s}-t$. Then obviously ${\cal N} = 0, 1, 2, \cdots,$ %take only non-negative integer values, i.e.
%\beq
%$2\k-q_{M_1}-q_{M_1+M_2}-t=0,1, \cdots$.
%\eeq
and the Fock states become %corresponding to the realization (\ref{su2-poly-boson}) are
\beqa
&&|\{q \}^{(1)}, \{q \}^{(2)}, n,\{l\}^{(1)},  \{l \}^{(2)},\k \ra =|\{q \}^{(1)}, n,\{l \}^{(1)} \ra \;
  |\{q \}^{(2)}, {\cal N}-n,\{l \}^{(2)} \ra \n
&& ~~~~~~=\frac{ \prod_{i=1}^{r}(a_i^{\dg})^{k_i(n+q_{r}+s_i^{(1)}-\frac{1}{k_i^2} )}
\prod_{i=r+1}^{r+s}(a_i^{\dg})^{k_i(2\k-q_{r}-t+s_{i}^{(2)}-\frac{1}{k_i^2}-n )} |0 \ra}
{\sqrt{\prod_{i=1}^{r}[k_i (n+q_{r}+s_i^{(1)}-\frac{1}{k_i^2} )]!
\prod_{i=r+1}^{r+s} [ k_i(2\k-q_{r}-t+s_{i}^{(2)}-\frac{1}{k_i^2}-n )] !}}, \n
%\n
%&&~~~~~~~~~~ \times { \prod_{i=M_1+1}^{M_1+M_2}(a_i^{\dg})^{k_i\left(2\k-q_{M_1}-t+s_{i}^{(2)}-\frac{1}{k_i^2}-n \right)} 
%\over \sqrt {\prod_{i=M_1+1}^{M_1+M_2} \left[ k_i\left(2\k-q_{M_1}-t+s_{i}^{(2)}-\frac{1}{k_i^2}-n \right)\right] !}} |0 \ra, \n 
%\n
%&&~~~~~~n=0,1, \,\cdots  ,\, 2\k-q_{M_1}-q_{M_1+M_2}-t,
\eeqa
where $n=0,1, \,\cdots  ,\, {\cal N}$. 
% noting that  $2\k-q_{M_1}-q_{M_1+M_2}-t$ is always less than or equal to 
%$2\k-q_{M_1}-t+s_i^{(2)}-\frac{1}{k_{i}^2}$ for $i=M_1+1, \cdots , M_1+M_2$ (since $ s_i^{(2)}\geq q_i-q_{M_1+M_2}$).
This gives us the ${\cal N}+1$ dimensional irreducible representation of ({\ref{su2-poly}),
\beqa
&&{\cal P}_0|\{q \}^{(1)}, \{q \}^{(2)}, n,\{l\}^{(1)},  \{l \}^{(2)},\k \ra  \n
&&~~~~~~= \left(-\k+q_{r}+n+\frac{1}{r}\sum_{j=1}^{r}s_i^{(1)} \right)  
  |\{q \}^{(1)}, \{q \}^{(2)}, n,\{l\}^{(1)} ,  \{l \}^{(2)},\k \ra , \n
&&{\cal P}_+|\{q \}^{(1)}, \{q \}^{(2)}, n,\{l\}^{(1)} , \{l \}^{(2)},\k \ra  \n
&&~~~~~~= \pd_{j=r+1}^{r+s}\pd_{i=1}^{k_j}\left( 2\k-n-q_{r}-t+s_j^{(2)}- \frac{(i-1)k_j+1}{k_j^2} 
  \right)^\frac{1}{2} \n
&&~~~~~~~~ \times \pd_{j=1}^{r}\pd_{i=1}^{k_j}\left(n+q_{r}+s_{j}^{(1)}+\frac{ik_j-1}{k_j^2} \right)^\frac{1}{2} %\n
%&&~~~~~~~~ \times 
|\{q \}^{(1)}, \{q \}^{(2)}, n+1,\{l\}^{(1)} , \{l \}^{(2)},\k \ra ,\n
&&{\cal P}_-|\{q \}^{(1)}, \{q \}^{(2)}, n,\{l\}^{(1)},  \{l \}^{(2)} ,\k \ra \n
&&~~~~~~= \pd_{j=r+1}^{r+s}\pd_{i=1}^{k_j}\left( 2\k-n-q_{r}-t+s_j^{(2)}+\frac{ik_j-1}{k_j^2}\right)^\frac{1}{2}\n
&&~~~~~~~~ \times \pd_{j=1}^{r}\pd_{i=1}^{k_j}\left( n+q_{r}+s_j^{(1)} -\frac{(i-1)k_j+1}{k_j^2}\right)^\frac{1}{2} %\n
%&&~~~~~~~~ \times 
  |\{q \}^{(1)}, \{q \}^{(2)}, n-1,\{l\}^{(1)},  \{l \}^{(2)},\k \ra . \n
\label{su2-poly-rep}
\eeqa

By using the standard Fock-Bargmann correspondence,
%\beq
%a_i^{\dg} \longrightarrow z_i , \hspace{1cm} a_i \longrightarrow \frac{d}{dz_i},
%\hspace{1cm} |n_i\ra \longrightarrow \frac{z_i^{n_i}}{\sqrt{n_i!}},
%\eeq
%we can make the following association
%\beqa
%&&|\{q \}^{(1)},\{q \}^{(2)},n,\{l\}^{(1)}, \{l\}^{(2)},\k \ra  \longrightarrow \n 
%&&\frac{\pd_{i=1}^{M_1}z_i^{k_i(n+q_{M_1}+s_i^{(1)}-\frac{1}{k_i^2})}
%\pd_{i=M_1+1}^{M_1+M_2}z_i^{k_i(2\k-q_{M_1}-n+s_i^{(2)}-\frac{1}{k_i^2})}}
%{\sqrt{\pd_{i=1}^{M_1}\left( k_i(n+q_{M_1}+s_i^{(1)}-\frac{1}{k_i^2})\right)!}
%\sqrt{\pd_{i=M_1+1}^{M_1+M_2}\left( k_i(2\k-q_{M_1}-t-n+s_i^{(2)}-\frac{1}{k_i^2}\right)!}}.\n
%\eeqa
%Now since $\{l \}^{(1,2)},\{q \}^{(1,2)}$ are constants, 
we can map the Fock states
$|\{q \}^{(1)},\{q\}^{(2)},n,\{l\}^{(1)}, \{l\}^{(2)}, \k\ra $ to the monomials in $z = \pd_{i=1}^{r}z_i^{k_i} /
\prod_{i=r+1}^{r+s}z_i^{k_i}$,
$$
\Psi (z)=\frac{z^n}{\sqrt{\pd_{i=1}^{r}[ k_i(n+q_{r}+s_i^{(1)}-\frac{1}{k_i^2})]!
\pd_{i=r+1}^{r+s}[ k_i(2\k-q_{r}-t+s_i^{(2)}-\frac{1}{k_i^2}-n)]!}}, 
$$
%\sqrt{\pd_{i=M_1+1}^{M_1+M_2}\left( k_i(2\k-q_{M_1}-t-n+s_i^{(2)}-\frac{1}{k_i^2}\right)!}},
\beq
n=0,1,\,\cdots,\, {\cal N}.
\eeq
The corresponding single-variable differential operator realization of (\ref{su2-poly}) is
\beqa
{\cal P}_0 &=& z\frac{d}{dz}-\k+ q_{r} +\frac{1}{r}\sum_{j=1}^{r}s_j^{(1)}, \n
{\cal P}_+ &=& z\frac{\pd_{p=r+1}^{r+s}\left(\sqrt{k_p}\right)^{k_p}}{\pd_{p=1}^{r}\left(\sqrt{k_p}\right)^{k_p}}
\pd_{j=r+1}^{r+s}\pd_{i=1}^{k_j}\left( -z\dz-q_{r}-t+2\k+s_j^{(2)}-\frac{(i-1)k_j+1}{k_j^2}\right), \n
{\cal P}_- &=& z^{-1}\frac{\pd_{p=1}^{r}\left(\sqrt{k_p}\right)^{k_p}}{\pd_{p=r+1}^{r+s}
  \left(\sqrt{k_p}\right)^{k_p}}\pd_{j=1}^{r}\pd_{i=1}^{k_j}
   \left(z\dz+q_{r}+s_j^{(1)}- \frac{(i-1)k_j+1}{k_j^2} \right).\label{su2-poly-d}
\eeqa
These differential operators form the same $({\cal N}+1)$ dimensional representations in the space of monomials
as those realized by (\ref{su2-poly-boson}) in the corresponding Fock space. We remark that because $\pd_{j=1}^{r}\pd_{i=1}^{k_j}
   \left(q_{r}+s_j^{(1)} - \frac{(i-1)k_j+1}{k_j^2} \right)\equiv 0$ for all the allowed $q_{r}$ values (noting 
$s_{r}^{(1)}=0$) there is no $z^{-1}$ term in ${\cal P}_-$ above and thus the differential operator 
expressions (\ref{su2-poly-d}) are non-singular.

%%%%%%%%%%%%%%%%%%%%%%%%%%%%%%%%%%%%%%%%%%%%%%%%%%%%%%%%%%%%%%%%%%%%%%%%%%%%%%%%%%%%%%%%%%%%%%%%%%%%%%%%%%%%%%%%%%%%%%%%%%%%%%%%%%%%%%%%%%%%%%%%

\sect{Exact solution of the multi-mode boson systems}

We now use the differential operator realization (\ref{su2-poly-d})
to exactly solve the multi-mode boson Hamiltonian (\ref{bosonH1}).
%\beq H=\sum_{i,j}^2 w_{ij}N_iN_j + \sum_i^2 w_iN_i+ g\left( a_1^{\dagger s}a_2^r +a_1^sa_2^{\dagger r} \right) \eeq

By (\ref{jordan-schwinger}) and  the multi-boson realization (\ref{su2-poly-boson}),
%identifying $k_i$ with $m_i$,  $M_1$ with $r$ and $M_2$ with $s$, 
we may express the Hamiltonian (\ref{bosonH1}) 
in terms of the generators of the polynomial algebra (\ref{su2-poly}), % with Casimir values (\ref{Ci-values}),
\beq
H=\sum_i^{r+s} w_iN_i+\sum_{i \le j}^{r+s} w_{ij}N_iN_j + g\pd_{i=1}^{r+s}\left(\sqrt{k_i}\right)^{k_i}\left( {\cal P}_+ + {\cal P}_- \right)\label{bosonH2}
\eeq
with the number operators having the following expressions in ${\cal P}_0$ and $\cal L$
\beq 
N_i =\left\{
\begin{array}{ll}
 k_i\left({\cal K}+ {\cal P}_0+ \sum_{\mu=i}^{r-1}{\cal L}_\mu-\frac{1}{r}\sum_{\mu=1}^{r-1}\mu{\cal L}_\mu \right) 
  -\frac{1}{k_i} & {\rm for}~i =1, \cdots , r\\
k_i\left({\cal K}-{\cal P}_0 +\sum_{\mu=i}^{r+s-1}{\cal L}_\mu-\frac{1}{s}\sum_{\mu=r+1}^{r+s-1}\mu{\cal L}_\mu \right) 
 -\frac{1}{k_i}  & {\rm for}~ i = r+1, \cdots , r+s
\end{array}
\right..
\eeq 
In deriving the above expressions for $N_i$, we have used the relationships between ${\cal Q}_0^{(1, 2)}$ and $Q_0^{(i)}$ similar to
those given in (\ref{QL-relations}).
Keep in mind that $\{{\cal P}_{\pm,0}\}$ in (\ref{bosonH2}) as realized by (\ref{su2-poly-boson}) (and (\ref{jordan-schwinger})) form
the ${\cal N}+1$ dimensional representation of the polynomial algebra (\ref{su2-poly}). 
%where ${\cal N}\equiv 2\k-q_r-q_{r+s}-t~(=0,1,\cdots)$. 
This representation is also
realized by the differential operators (\ref{su2-poly-d}) acting on the ${\cal N}+1$  dimensional space of polynomials with basis
 $\left\{1,z,z^2,...,z^{\cal N} \right \} $. We can thus equivalently represent (\ref{bosonH2}) (i.e. (\ref{bosonH1}))
as the single-variable differential operator of order ${\cal M} \equiv$ max$\{\sum_{i=1}^r k_i,\sum_{i=r+1}^{r+s}k_i, 2\}$,
\beqa
H &=&\sum_i^{r+s} w_iN_i+\sum_{i,j}^{r+s} w_{ij}N_iN_j\n
 & &+gz \pd_{j=r+1}^{r+s} \pd_{i=1}^{k_j} k_j\left(-z\dz -q_{r}-t+2\k+s_j^{(2)}-\frac{(i-1)k_j+1}{k_j^2} \right)\n
 & &+gz^{-1}\pd_{j=1}^{r} \pd_{i=1}^{k_j}k_j\left(z\dz+q_r+s_j^{(1)}- \frac{(i-1)k_j+1}{k_j^2} \right)\label{differentialH}
\eeqa
with
\beq 
N_i = \left\{
\begin{array}{ll}
k_i(z\frac{d}{dz}+ q_r+s_i^{(1)} ) -\frac{1}{k_i} & {\rm for}~i = 1, \cdots , r\\
 k_i(2\k-z\frac{d}{dz}-q_r-t+s_i^{(2)}) -\frac{1}{k_i} & {\rm for}~i = r+1, \cdots, r+s
\end{array}
\right..
\eeq

We will now solve for the Hamiltonian equation
\beq
H\psi(z)=E\,\psi(z)\label{hamilton-eqn}
\eeq
 by using the Functional Bethe Ansatz method,
where $\psi(z)$ is the eigenfunction and $E$ is the corresponding eigenvalue. It is easy to verify
\beqa
 Hz^n&=&z^{n+1}\,g\pd_{j=r+1}^{r+s}\pd_{i=1}^{k_j}k_j \left( -n-q_r-t+2\k+s_j^{(2)}- \frac{(i-1)k_j+1}{k_j^2}\right) \n 
   &&+~{\rm lower~order~terms},~~~~~~~~~  n\in {\bf Z}_+.\label{hzm=hzm+1}
\eeqa
This means that the differential operator ({\ref{differentialH}) is not exactly solvable.
However, it is quasi exactly solvable, since it has an invariant polynomial subspace of degree ${\cal N}+1$:
\beq
H{\cal V} \subseteq  {\cal V}, ~~~~~{\cal V}= {\rm span} \{1,z,...,z^{\cal N}\},~~~~~{\rm dim}{\cal V}={\cal N}+1.
\eeq
This is easily seen from the fact that when $n={\cal N}$, the first term on the r.h.s. of (\ref{hzm=hzm+1}) becomes
$z^{{\cal N}+1}\,g\pd_{j=r+1}^{r+s}\pd_{i=1}^{k_j} k_j\left( q_{r+s}+s_j^{(2)}-\frac{(i-1)k_j+1}{k_j^2}\right)$ 
which vanishes identically for all the allowed $q_{r+s}$ values (noting $s_{r+s}^{(2)}=0$).

As (\ref{differentialH}) is a quasi exactly solvable differential operator preserving ${\cal V}$,
up to an overall factor, its eigenfunctions have the form,
\beq
\psi(z)= \prod_{i=1}^{ \cal N}\left(z-\alpha_i\right),\label{w-function}
\eeq
where  $\{\alpha_i\,|\,i=1,2,\cdots,{ \cal N}\}$ are roots of the polynomial
which will be specified later by the associated Bethe ansatz equations (\ref{bethe-ansatz-eqn}) below.
We can rewrite the Hamiltonian (\ref{differentialH}) as
\beq
 H = \sum_{i=1}^{\cal M} P_i(z)\left(\frac{d}{dz} \right)^i + P_0(z),\label{expansionH}
\eeq
%% where
%%\beqa
%%P_0(z)&=&
%%\eeqa
%%and $P_i(z)$ are the coefficients in front of $d^i/dz^i$ in the expansion of %%(\ref{differentialH}) (see the Appendix),
%%\beqa
%%P_i(z) &=& 
%%\eeqa
%%In the above expression, % $\delta_{i\leq t}$ equals to 1 when $i\leq t$ and to 0 %%otherwise, and
%%\beqa
%%A_i &=&  
%%\eeqa
where $P_0(z)$ and $P_i(z)$ are polynomials in $z$ resulted from the expansion of the products in (\ref{differentialH}). 

Dividing the Hamiltonian equation $H\psi= E\psi $ over by $\psi$ gives us
\beq
E= \frac{H\psi}{\psi} = \sum_{i=1}^{\cal M}P_i(z)i!\sum_{m_1<m_2<...<m_i}^{\cal N}
\frac{1}{(z-\alpha_{m_1})...(z-\alpha_{m_i})}  +P_0(z).\label{e=hpsi/psi}
\eeq
The l.h.s. of (\ref{e=hpsi/psi}) is a constant, while the r.h.s is a meromorphic function in $z$ with at most simple poles.
For them to be equal, we need to eliminate all singularities on the r.h.s of (\ref{e=hpsi/psi}). We may achieve this by demanding
that the residues of the simple poles, $z=\alpha_i, i=1,2,..., { \cal N}$ should all vanish. This leads to the Bethe ansatz
equations for the roots $\{\alpha_i \}$ :
\beqa 
&&\sum_{i=2}^{\cal M}\;\sum_{m_1<m_2<...<m_{i-1} \ne p}^{ \cal N}\frac{P_i(\alpha_p)i!}{(\alpha_p-\alpha_{m_1})\cdots
(\alpha_p-\alpha_{m_{i-1}})}+P_1(\alpha_{p})=0,  ~~~~~p=1,2,\,\cdots,\,{ \cal N}.  \n
 \label{bethe-ansatz-eqn}
\eeqa
The wavefunction $\psi(z)$ (\ref{w-function}) becomes the eigenfunction of $H$ (\ref{differentialH}) in the space ${\cal V}$
provided that the roots $\{\alpha_i\}$ of the polynomial $\psi(z)$ (\ref{w-function}) are the solutions of (\ref{bethe-ansatz-eqn}).

Let us remark that the Bethe ansatz equation (\ref{bethe-ansatz-eqn}) is the necessary and sufficient condition for
the r.h.s. of (\ref{e=hpsi/psi}) to be independent of $z$. This is because when (\ref{bethe-ansatz-eqn}) is satisfied
the r.h.s. of (\ref{e=hpsi/psi}) is analytic everywhere in the complex plane (including points at infinity) and thus must be
a constant by the Liouville theorem.

To get the corresponding eigenvalue $E$, we consider the leading order expansion of $\psi(z)$,	
$$\psi(z)= z^{ \cal N}- z^{{ \cal N}-1}\sum_{i=1}^{ \cal N}\alpha_{i} +\cdots. $$
It is easy to show that ${\cal P}_{\pm, 0}\psi(z)$ have the  expansions,
\beqa
{\cal P}_+\psi &=&  -z^{ \cal N} \frac{\pd_{p=r+1}^{r+s}\left(\sqrt{k_p}\right)^{k_p}}
 {\pd_{p=1}^{r}\left(\sqrt{k_p}\right)^{k_p}}\n
& &\times \left[\pd_{j=r+1}^{r+s}\pd_{i=1}^{k_j}\left( q_{r+s}+1+s_j^{(2)}-\frac{(i-1)k_j+1}{k_j^2} \right)\right]  
  \sum_{i=1}^{ \cal N}\alpha_i +\cdots,  \n
{\cal P}_-\psi &=& z^{{ \cal N}-1}\frac{\pd_{p=1}^{r}(\sqrt{k_p})^{k_p}}{\pd_{p=r+1}^{r+s}(\sqrt{k_p})^{k_p}} 
\pd_{j=1}^{r}\left({\cal N}+q_{r}+s_j^{(1)}- \frac{(i-1)k_j+1}{k_j^2} \right)+ \cdots,  \n
{\cal P}_0\psi &=& z^{{ \cal N}}\left( \k-q_{r+s}-\frac{1}{s}\sum_{j=r+1}^{r+s}s_j^{(2)}\right)+  \cdots.
\eeqa
%The first term on the right hand side of the first equation vanishes identically
%because $\pd_{j=1}^{r}\left( q_2-\frac{(j-1)r+1}{r^2} \right)\equiv 0$ for all the allowed $q_2$ values.
%The Hamiltonian equation therefore has the following leading term expansion
%\beqa
%H\psi &= & z^M \left[w_{11}\left( s(2l- q_2 )- \frac{1}{s}\right)^2+w_{22}\left( rq_2 -\frac{1}{r} \right)^2\right.\n
%&&+ (w_{12}+w_{21})\left( s(2l- q_2 )- \frac{1}{s}\right)\left( rq_2  -\frac{1}{r} \right)  \n
%&&+w_{1}\left(s(2l- q_2) -\frac{1}{s}  \right)+w_2\left(rq_2  -\frac{1}{r} \right)  \n
%&&\left.-g \left(\pd_{j=1}^{r} r\left(  q_2+1-\frac{(j-1)r+1}{r^2}\right)\right)
%  \sum_{i=1}^M\alpha_i  \right]  +\cdots \n
%&=& Ez^M +\cdots.
%\eeqa
Substituting these expressions into the Hamiltonian equation (\ref{hamilton-eqn}) and equating the $z^{ \cal N}$ terms, we arrive at
\beqa
E &=& \sum_{i=1}^r w_{ii}\left( k_i({\cal N}+q_{r}+s_i^{(1)})-\frac{1}{k_i} \right)^2
+\sum_{i=r+1}^{r+s}w_{ii}\left( k_i(q_{r+s}+s_i^{(2)}) -\frac{1}{k_i} \right)^2\n
&&+ \sum_{j=r+1}^{r+s}\sum_{i=1}^{r}w_{ij}\left( k_i({\cal N}+q_{r}+s_i^{(1)})-\frac{1}{k_i} \right) 
\left( k_{j}(q_{r+s} +s_j^{(2)} )- \frac{1}{k_{j}}\right) \n
&&+ \sum_{i < j=2}^r w_{ij}\left( k_i({\cal N}+q_{r}+s_i^{(1)})-\frac{1}{k_i} \right)
\left( k_j({\cal N}+q_{r}+s_j^{(1)})-\frac{1}{k_j} \right) \n 
&&+ \sum_{i < j=r+2}^{r+s}w_{ij}\left( k_i(q_{r+s}+s_i^{(2)}) -\frac{1}{k_i} \right)
\left( k_j(q_{r+s}+s_j^{(2)}) -\frac{1}{k_j} \right) \n
&&+ \sum_{i=1}^r w_{i}\left(k_i({\cal N}+q_{r}+s_i^{(1)})-\frac{1}{k_i} \right)
+\sum_{i=r+1}^{r+s}w_i\left( k_i(q_{r+s}+s_i^{(2)}) -\frac{1}{k_i} \right)  \n
&&-g\left[\pd_{j=r+1}^{r+s}\pd_{i=1}^{k_j}k_j\left( q_{r+s}+1+s_j^{(2)}-\frac{(i-1)k_j+1}{k_j^2} \right)\right]
  \sum_{i=1}^{ \cal N}\alpha_i ,\label{energy-generalH}
\eeqa
where $\{\alpha_i\}$ satisfy the Bethe ansatz equations (\ref{bethe-ansatz-eqn}). This gives
the eigenvalue of the 2-mode boson Hamiltonian (\ref{bosonH1}) with the corresponding eigenfunction $\psi(z)$ (\ref{w-function}).

%%%%%%%%%%%%%%%%%%%%%%%%%%%%%%%%%%%%%%%%%%%%%%%%%%%%%%%%%%%%%%%%%%%%%%%%%%%%%%%%%%%%%%%%%%%%%%%%%%%%%%%%%%%%%%%%%%%%%
\sect{ Explicit examples corresponding to BECs}

In this section we give explicit results on the Bethe ansatz equations and energy eigenvalues
of the Hamiltonian (\ref{bosonH1}) for the special cases of $r+s \le 4$. 
These models arise in the description of Josephson tunneling effects and atom-molecule conversion processes 
in the context of BECs. Our approach provides a unified treatment of their exact solutions. \\

\noindent \textbf{A. $r=2, s=1, ~~k_1= k_2= k_3=1$}\\

The Hamiltonian is
\beq
H=\sum_i^3 w_iN_i+\sum_{i\leq j}^3 w_{ij}N_iN_j +  g\left( a_1^{\dg }a_2^{\dg}a_3 +a_1a_2a_3^{\dagger } \right).\label{h11}
\eeq
This is the so-called hetero-atom-molecule BEC model and has been solved in \cite{Andreev02,Links03}
via the Algebraic Bethe Ansatz method. Here for completeness
we present the exact solution derived from our approach without providing any details.
%  From the general results in the preceding section, in this case, we have $q_2=1=q_3$, $t=\frac{s_1^{(1)}}{2}=\frac{l_1}{2}$
%and ${\cal N}=2\k-2-\frac{l_1}{2}$. Thus (\ref{h11}) takes the form,
%\beq
%H= P_2(z) \frac{d^2}{dz^2}+ P_1(z) \dz +P_0(z),
%\eeq
%where
%\beqa
%P_2(z)&=& gz+A_{11}z^2, \n
%P_1(z)&=& +B_{11}z +g \left(( l_1+1  -{z}^{2}\right),\n
%P_0(z) &=& g\left(2\k-2-\frac{l_1}{2}\right) z+D_{11}
%\eeqa
%with
%\beqa
%A_{11} &=& w_{22}+w_{11}+w_{33}+w_{12}-w_{13}-w_{23},\n
%B_{11} &=& w_1 -w_{{3}}+w_{{2}}+w_{{22}}+w_{{11}} \left( 2l_1+1 \right) 
%  +w_{{33}} \left( 5+l_1-4\k \right)\n
%& &  + w_{{13}} \left( 2\k-3-\frac{3}{2}l_1 \right) +w_{{12}} \left( l_1+1 \right) 
%  +w_{{23}} \left( 2\k-3-\frac{l_1}{2} \right),   \n
%D_{11} &=& w_{1}l_1+w_{33}\left( 2\k-2-\frac{l_1}{2} \right)^{2}+(w_{3}+w_{13}l_1)\left( 2\k-2-\frac{l_1}{2} \right)
% +w_{11} l_1^2.
%\eeqa

The Bethe ansatz equations are given by
\beqa
\sum_{i \ne p}^{{\cal N}} {2\over \a_i - \a_p}=  {B_{11}\a_p +g \left[\left( l_1+1 \right) 
-{\a_p}^{2}\right] \over g\a_p+A_{11}\a_p^2 },~~~~p=1,2,\cdots, {\cal N}
\eeqa
with ${\cal N}\equiv 2\k-2-\frac{l_1}{2}$ and the energy eigenvalues are
\beqa
E&=&w_{11}\left({\cal N}+l_1 \right)^2+w_{22} {\cal N}^2
  +w_1\left({\cal N}+l_1 \right)+w_2{\cal N}+ w_{12}{\cal N}  \left({\cal N}+l_1 \right)  -g \sum_{i=1}^{{\cal N}} \a_i.\n
\eeqa
Here 
\beqa
A_{11} &=& w_{22}+w_{11}+w_{33}+w_{12}-w_{13}-w_{23},\n
B_{11} &=& w_1 -w_{{3}}+w_{{2}}+w_{{22}}+w_{{11}} \left( 2l_1+1 \right) +w_{{33}} \left( 5+l_1-4\k \right).
\eeqa

\vskip.2in
\noindent\textbf{B. $r=2, s=1,~~ k_1= k_2=1, k_3=2$}\\

The Hamiltonian is
\beq
H=\sum_i^3 w_iN_i+\sum_{i\leq j}^3 w_{ij}N_iN_j +  g\left( a_1^{\dagger }a_2^{\dg} a_3^2 +a_1a_2a_3^{\dagger 2} \right).\label{h21}
\eeq
This yields a model of three-mode atomic-molecular BECs which has not been exactly solved previously.
Specializing the general results in the preceding section to this case,
we have $q_2=1$, $q_3={1 \over 4}$ or $ 3\over 4$, $t=\frac{s_1^{(1)}}{2}=\frac{l_1}{2}$
and ${\cal N}=2\k-1-q_3-\frac{l_1}{2}$.
The differential operator representation of the Hamiltonian (\ref{h21}) is thus
\beq
H= P_2(z) \frac{d^2}{dz^2}+ P_1(z) \dz +P_0(z)
\eeq
where
\beqa
P_2(z)&=& 4gz^{3}+ A_{21} {z}^{2}+gz,\n
P_1(z)&=&  B_{21}{z}^{2}+ D_{21}z+g ( l_1+1 ),\n
 P_0(z)&=& F_{21}z 
\eeqa
with
\beqa
A_{21} &=&  w_{11}+w_{22}-2w_{23}+4w_{33}-2w_{13}+w_{12} , \n
B_{21} &=& 4g \left( 4+l_1-4\k \right), \n
D_{21} &=&  w_{2}+w_{1}+w_{22}+w_{12} \left(l_1+1 \right) -2w_{3}+w_{33}
 \left( 14+4l_1-16\k \right)\n
& & +w_{11} \left( 2l_1+1 \right)
 +w_{13} \left( 4\k-{9 \over 2}-3l_1 \right) +w_{23}\left( 4\k-{9 \over 2}-l_1 \right),\n
F_{21} &=& g \left( (4\k-2-l_1)(4\k-4-l_1)+\frac34 \right) ,\n 
G_{21} &=& w_{{1}}l_1+w_{{3}} \left( 4\k-\frac52-l_1 \right) +w_{{13}}l_1
 \left( 4\k-l_1-\frac52 \right) \n 
&& +w_{{11}}l_1^{2}+w_{{33}} \left( (4\k-l_1-\frac52 \right)^2. 
\eeqa
The Bethe ansatz equations are given by
\beq
\sum_{i \ne p}^{ {\cal N} } {2\over \a_i - \a_p}=  {  B_{21}\a_p^{2}+ D_{21}\a_p+g ( l_1+1 ) 
\over  4g\a_p^{3}+ A_{21} \a_p^{2}+g\a_p },~~~~p=1,2,\cdots, {\cal N}
\eeq
and the energy eigenvalues are given by
\beqa
E &=& w_{11}\left({\cal N}+l_1\right)^2+ w_{22}{\cal N}^2  +2w_{33}\left(q_3 - {1 \over 4} \right)^2
+w_{12}{\cal N}\left({\cal N}+l_1\right)\n
& & +2\left(q_3-\frac14\right)\left[w_{13}\left({\cal N}+l_1\right) +w_{23}{\cal N}+w_3\right] 
+w_1\left({\cal N}+l_1\right)+w_2{\cal N}\n
& &-4g\left(q_3+\frac14\right)\left(q_3+\frac34\right)  \sum_{i=1}^{{\cal N}}\a_i,
\eeqa
where ${\cal N}=2\k-1-q_3-\frac{l_1}{2}$.

\vskip.2in
\noindent \textbf{C. $r=s=2,~~ k_1=k_2=k_3=k_4=1$}\\

The Hamiltonian is
\beq
H=\sum_i^4 w_iN_i+\sum_{i\leq j}^4 w_{ij}N_iN_j +g\left( a_1^{\dagger }a_2^{\dg}a_3a_4 +a_1a_2a_3^{\dg}a_4^{\dg} \right).\label{h22}
\eeq
This gives a model of four-mode atom-molecule BECs. This model has not been exactly solved previously.
Applying the results in the preceding section gives $q_2=1=q_4$, $t=\frac12(s_1^{(1)}+s_3^{(2)})=\frac12(l_1+l_3)$ and
${\cal N}=2\k-2-\frac12(l_1+l_3)$. The differential operator representation of the Hamiltonian (\ref{h22}) thus reads
\beq
H= P_2(z) \frac{d^2}{dz^2}+ P_1(z) \dz +P_0(z),
\eeq
where
\beqa
P_2(z)&=& g(z^3+z) +A_{22} z^2,\n
P_1(z)&=&  B_{22}z^2 + D_{22}z + g(l_1+1),\n
P_0(z)&=& g{\cal N}\left({\cal N}+l_3\right) z + G_{22}
\eeqa
with
\beqa
A_{22} &=&w_{{11}}+w_{{33}}+w_{{22}}-w_{{24}}-w_{{13}}+w_{{12}}-w_{{23}}+w_{{44}}-w_{{14}}+w_{{34}},\n
B_{22} &=& g \left( l_1+5-4\k \right),\n
D_{22}&=& w_1+w_2-w_{{4}}-w_{{3}}+w_{{22}}+w_{{33}} \left( 1-2l_{{3}}-2{\cal N} \right) +w_{{11}} \left( 2l_{{1}}+1 \right)\n
& &+w_{{12}} \left( 1+l_{{1}} \right)+w_{{23}} \left({\cal N}+l_3-1\right)+w_{{13}} \left({\cal N}+l_3-l_1-1 \right)\n
& &   +w_{{14}} \left( {\cal N}-l_1-1 \right) +w_{{44}} \left( 1-2{\cal N} \right) +w_{{34}} 
\left( 1-l_3-2{\cal N} \right) +w_{{24}} {\cal N},\n
G_{22}&=& w_{{1}}l_{{1}} +w_{{3}} \left( {\cal N}+l_3\right) +w_{{4}} {\cal N}+w_{{33}} 
\left({\cal N}+l_3\right)^2 +w_{{34}} {\cal N}\left( {\cal N}+l_3 \right)\n  
& & +w_{{11}}{l_{{1}}}^{2}+w_{{14}}l_1{\cal N}+w_{{44}}{\cal N}^2  +w_{{13}} l_1\left( {\cal N}+l_3\right), 
\eeqa
where (and below) ${\cal N}=2\k-2-\frac12(l_1+l_3)$.

The Bethe ansatz equations read
\beq
\sum_{i \ne p}^{ {\cal N} } {2\over \a_i - \a_p}=  { B_{22}\a_p^2 + D_{22}\a_p + g(l_1+1) 
\over g(\a_p^3+\a_p) +A_{22} \a_p^2 },~~~~ p=1,2,\cdots, {\cal N}
\eeq
and the energy eigenvalues are
\beqa
E &=& w_{11}\left({\cal N}+l_1\right)^2+w_{22}{\cal N}^2+w_{33}\,l_3^2 +w_{12}{\cal N}
\left({\cal N}+l_1\right)+w_{13}l_3\left({\cal N}+l_1\right)\n
& &+\left(w_{23}l_3+w_2\right){\cal N}  +w_1\left({\cal N}+l_1\right)+w_3\,l_3 - g (l_3+1) \sum_{i=1}^{ {\cal N}} \a_i.
\eeqa

\sect{Discussions}

We have introduced the higher order polynomial algebra (\ref{su2-poly}) and identified it as the dynamical symmetry algebra  
of the multi-boson system (\ref{bosonH1}). We have constructed its unitary irreducible representation 
(\ref{su2-poly-rep}) and the corresponding
single-variable differential realization (\ref{su2-poly-d}). 
We have then used the differential realization to rewrite the Hamiltonian (\ref{bosonH1})
as a QES differential operator (\ref{differentialH}) acting on the finite dimensional monomial space, 
thus providing an algebraization of the higher order Hamiltonian differential equation (\ref{hamilton-eqn}). 
Exact eigenfunctions and eigenvalues of the Hamiltonian (\ref{bosonH1}) have been
found by employing the Functional Bethe Ansatz technique.
As examples, we have provided explicit expressions for the BEC models which correspond to the $r+s \le 4$ cases of
(\ref{bosonH1}). % and established the spectral correspondence of these specific models with QES Schr\"odinger potentials. 

It would be interesting to adapt the techniques of this paper to obtain exact solutions of the generalized 
Lipkin-Meshkov-Glick and Tavis-Cummings type model defined by  
\beqa 
H&=& \sum_{i=1}^{r}w_i N_i + \Omega J_0^s +g \left( J_+^k a_{1}^{k_{1}}\cdots a_{r}^{k_{r}}
 + J_-^k a_1^{\dagger k_1}\cdots a_r^{\dagger k_r}\right), 
\eeqa 
where $J_{\pm,0}$ are the generators of the $su(2)$ algebra. Also we plan to explore the q-boson counterparts of the 
multi-boson systems and  their underlying q-deformed polynomial algebraic structures.
Lastly, it is of interest to further explore the connection between the higher order ODEs and integrable models 
(i.e. the ODE/IM correspondence) along the line of \cite{Dorey99}. Results in those directions will be presented elsewhere.

%\vskip.2in
%\noindent {\bf Acknowledgments:} This work was supported by the Australian Research Council.

\bebb{99}

\bbit{Yuan10}
Y.-H. Lee, W.-L. Yang and Y.-Z. Zhang, J. Phys. A: Math. Theor. {\bf 43}, 185204 (2010).

\bbit{Smith90}
S.P. Smith, Trans. Amer. Math. Soc. {\bf 322}, 285 (1990).

\bbit{Higgs79}
P.W. Higgs, J. Phys. A: Math. Gen. {\bf 12}, 309 (1979).

\bbit{Rocek91}
M. Rocek, Phys. Lett. B {\bf 255}, 554 (1991).

\bbit{Schoutens91}
K Schoutens, A. Sevrin and P. Van Nieuwenhuizen, Comm. Math. Phys. {\bf 124}, 87 (1991);
  Phys. Lett. B {bf 255}, 549 (1991).

\bbit{Granovsky92}
Ya.I. Granovsky, A.S. Zhedanov and I.M. Lutzenko, Ann. Phys. (NY) {\bf 217}, 1 (1992).

\bbit{Letourneau95}
P. Letourneau and L. Vinet, Ann. Phys. (NY) {\bf 243}, 144 (1995).

\bbit{Quesne94}
C. Quesne, Phys. Lett. A {\bf 193}, 249 (1994); SIGMA {\bf 3}, 067 (2007).

\bbit{Bonatsos94}
D. Bonatsos, C. Daskaloyannis and K. Kokkotas, Phys. Rev. A {\bf 50}, 3700 (1994).

\bbit{Karassiov94}
V.P. Karassiov and A. Klimov, Phys. Lett. A {\bf 191}, 117 (1994).

\bbit{Karassiov92-00}
V.P. Karassiov, J, Sov. Laser Res. {\bf 13}, 188 (1992); J. Phys. A: Math. Gen. {\bf 27}, 153 (1994);
J. Russian Laser Res. {\bf 21}, 370 (2000).
%Rep. Math. Phys. {\bf 40}, 235 (1997).

\bbit{Karassiov02}
V.P. Karassiov, A.A. Gusev and S.I. Vinitsky, Phys. Lett. A {\bf 295}, 247 (2002).

\bbit{Klishevich01} S.M. Klishevich and M.S. Plyushchay, Nucl. Phys. B {\bf 606}, 583 (2001).
%; ibid {\bf 616}, 403 (2001).

\bbit{Zhedanov92}
A.S. Zhedanov, Mod. Phys. Lett. A {\bf 7}, 507 (1992).

\bbit{Kumar01}
V.S. Kumar, B.A. Bambah, and R. Jagannathan, J. Phys. A: Math. Gen. {\bf 34}, 8583 (2001);
Mod. Phys. Lett. A {\bf 17}, 1559 (2002).

%\bbit{Duncan07}
%M. Duncan, A. Foerster, J. Links, E. Mattei, N. Oelkers and A.P. Tonel, Nucl. Phys. B {\bf 3}, 227 (2007).

%\bbit{Abdesselam96}
%B. Abdesselam, J. Beckers, A. Chakrabarti and N. Debergh, J. Phys. A: Math. Gen.
%{\bf 29}, 3075 (1996).

\bbit{Beckers99}
J. Beckers, Y. Brihaye and N. Debergh, J. Phys. A: Math. Gen. {\bf 32}, 2791 (1999).

%\bbit{Bogoliubov02}
%N. Bogoliubov, J. Timonen, and M. Zvonarev, (2002), cond-mat/0201335.
\bbit{Debergh00}
N. Debergh, J. Phys. A: Math. Gen. {\bf 33}, 7109 (2000).

\bbit{Turbiner88}
A. Turbiner, Comm. Math. Phys. {\bf 118}, 467 (1988);
Quasi-exactly-solvable differential equations, 1994, hep-th/9409068.

\bbit{Ushveridze94}
A.G. Ushveridze, Quasi-exactly solvable models in quantum mechanics, Institute of Physics
Publishing, Bristol, 1994.

\bbit{Gonzarez93}
A. Gonz\'arez-L\'opez, N. Kamran and P. Olver, Comm. Math. Phys. {\bf 153}, 117 (1993).

%\bbit{Debergh03}
%J. Ndimubandi, N. Debergh and B. van den Bossche, Mod. Phys. Lett. A {\bf 18}, 1013 (2003).

%\bbit{Dunning08}
%C. Dunning, K. Hibberd and J. Links, J. Phys. A: Math. Theor. {\bf 41}, 315211 (2008).

%\bbit{Hu00}
%B. Hu and L.-M. Kuang, Phys. Rev. A {\bf 62}, 023610 (2000).

%\bbit{Jeugt95}
%J. Van der Jeugt and R. Jagannathan, J. Math. Phys. {\bf 36}, 4507 (1995).

%\bbit{Anderson95}
%M.H. Anderson, J.R. Ensher, M.R. Matthews, C.E. Wieman and E.A. Cornell, Science {\bf 269}, 198 (1995).

%\bbit{Anglin02}
%J.R. Anglin and W. Ketterle, Nature {\bf 416}, 211 (2002).

%\bbit{Zoller02}
%P. Zoller, Nature {\bf 417}, 493 (2002).

%\bbit{Donley02}
%E.A. Donley, N.R. Claussen, S.T. Thompson and C.E. Wieman, Nature {\bf 417}, 529 (2002).

\bbit{Alvarez02} G. \'Alvarez, F. Finkel, A. Gonz\'alez-L\'opez and M.A. Rodr\'iguez,
  J. Phys. A: Math. Gen. {\bf 35}, 8705 (2002).
  
\bbit{Alvarez95} G. \'Alvarez and R.F. \'Alvarez-Estrada, J. Phys. A: Math. Gen. {\bf 28},
  5767 (1995); ibid {\bf 34}, 10045 (2001).

\bbit{Andreev02}
V.A. Andreev and O.A. Ivanova, J. Phys. A: Math. Gen. {\bf 35}, 8587 (2002).

\bbit{Links03}
J. Links, H.-Q. Zhou, R.H. McKenzie and M.D. Gould, J. Phys. A: Math. Gen. {\bf 36}, R63 (2003).

\bbit{Wiegmann94}
P.B. Wiegmann and A.V. Zabrodin, Phys. Rev. Lett. {\bf 72}, 1890 (1994);
 Nucl. Phys.  B {\bf 451}, 699 (1995).

\bbit{Sasaki09}
R. Sasaki, W.-L. Yang and Y.-Z. Zhang, SIGMA {\bf 5}, 104 (2009).

\bbit{Sasaki01} R. Sasaki and K. Takasaki, J. Phys. A: Math. Gen. {\bf 34}, 9533 (2001).

%\bbit{Links06}
%J. Links and K.E. Hibberd, SIGMA {\bf 2}, Paper 095 (2006).

%\bbit{Sasaki01}
%R. Caseiro, J. Francoise and R. Sasaki, J. Math. Phys. {\bf 42}, 5329 (2001).

%\bbit{Zaslavskii90} O.B. Zaslavskii, Phys. Lett. A {\bf 149}, 365 (1990).

%\bbit{Ullate05}
%D. Gomez-Ullate, N. Kamran and R. Milson, J. Phys. A: Math. Gen. {\bf 38}, 2005
%(2005).

%\bbit{Vadeiko03}
%I.P. Vadeiko, G.P. Miroshnichenko, A.V. Rybin and J. Timonen, Phys. Rev. A
%{\bf 67}, 053808(2003).

%\bbit{Bogo93}
%N.M. Bogoliubov, R.K. Bullough and G.D. Pang, Phys. Rev. B {\bf 47}, ?? (1993).

%\bbit{Bortz06} 
%M. Bortz and S. Sergeev,  Eur. Phys. J. B {\bf 51}, 395 (2006).

\bbit{Dorey99}
P. Dorey and R Tateo, J. Phys. A: Math. Gen. {\bf 32}, L419 (1999). 

\eebb

\end{document}